\documentclass[journal]{IEEEtran}
\usepackage{cite}
\usepackage[pdftex]{graphicx}
\graphicspath{{Figs/}}
\usepackage{epsfig} % for postscript graphics files
\usepackage{mathptmx} % assumes new font selection scheme installed
\usepackage{times} % assumes new font selection scheme installed
\usepackage{amsmath} % assumes amsmath package installed
\usepackage{amssymb}  % assumes amsmath package installed

\usepackage{array}
\usepackage{multirow}
\usepackage{booktabs}
\usepackage{dblfloatfix}
\usepackage{gensymb}
\usepackage{xcolor}
\usepackage{url}

\begin{document}
\title{Phased Array Beamforming Methods for \\ Powering Biomedical Ultrasonic Implants
}

\author{Braeden~C.~Benedict,~\IEEEmembership{Student~Member,~IEEE,}
       Mohammad~Meraj~Ghanbari,~\IEEEmembership{Student~Member,~IEEE,}
        and~Rikky~Muller,~\IEEEmembership{Senior~Member,~IEEE}
\thanks{Research supported by the National Science Foundation Graduate Research Fellowship Program, the sponsors of the Berkeley Wireless Research Center, the Hellman Fellows Fund, and The Weill Neurohub.}% <-this % stops a space
\thanks{B.C. Benedict is with UC Berkeley/UC San Francisco Joint Program in Bioengineering, Berkeley, CA 94720 USA. (email: braeden@berkeley.edu)}%
\thanks{M.M. Ghanbari and R. Muller are with the Department of Electrical Engineering and Computer Sciences, University of California, Berkeley, Berkeley, CA 94720 USA. R. Muller is also with the Chan Zuckerberg Biohub, San Francisco, CA 94158 USA.}%
\thanks{Manuscript submitted August 7, 2022.}}%

\markboth{}%
{Shell \MakeLowercase{\textit{et al.}}: Bare Demo of IEEEtran.cls for IEEE Journals}

\maketitle

\begin{abstract}
Millimeter-scale implants using ultrasound for power and communication have been proposed for a range of deep-tissue applications, including neural recording and stimulation. However, published implementations have shown high sensitivity to misalignment with the external ultrasound transducer. Ultrasonic beamforming using a phased array to these implants can improve tolerance to misalignment, reduce implant volume, and allow multiple implants to be operated simultaneously in different locations. This paper details the design of a custom planar phased array ultrasound system, which is capable of steering and focusing ultrasound power within a 3D volume. Analysis and simulation is performed to determine the choice of array element pitch, with special attention given to maximizing the power available at the implant while meeting FDA limits for diagnostic ultrasound. Time reversal is proposed as a computationally simple approach to beamforming that is robust despite scattering and inhomogeneity of the acoustic medium. This technique is demonstrated both in active drive and pulse-echo modes, and it is experimentally compared with other beamforming techniques by measuring energy transfer efficiency. Simultaneous power delivery to multiple implants is also demonstrated.
\end{abstract}


\IEEEpeerreviewmaketitle

\section{Introduction}

\begin{figure}[t]
    \centering
    \includegraphics[width=1\linewidth]{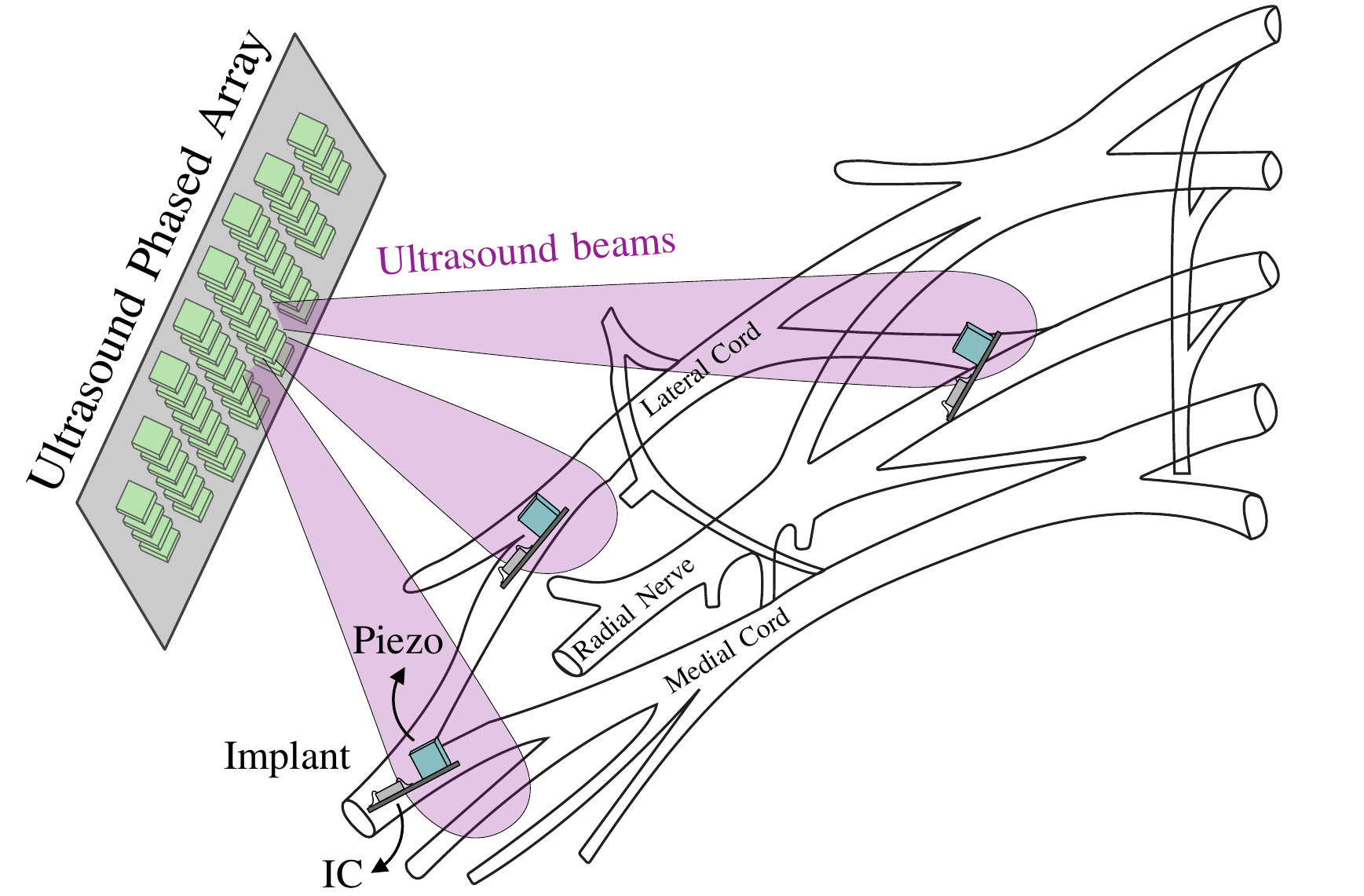}
    \caption[]{Concept of an ultrasound phased array powering multiple implants.}
    \label{fig:0}
\end{figure}

\IEEEPARstart{W}{ireless} millimeter-scale ultrasonic implants have been proposed to measure a range of physiological signals including neural activity \cite{ghanbari2019sub}, tissue oxygenation \cite{sonmezoglu2021monitoring}, and temperature \cite{shi2020monolithically}. They have also been proposed for applications such as electrical neural stimulation \cite{piech2020wireless, charthad2018mmsized}, optogenetic stimulation \cite{charthad2018mmsized}, and photodynamic tumor therapy \cite{kim2019implantable}. These miniaturized implants result in minimal tissue displacement and allow for untethered operation. While most implants use electromagnetic (EM) waves for wireless power and communication, ultrasound (US) has emerged as a promising alternative for deep-tissue implants. When compared with EM, US offers efficient propagation in tissue and a relatively small wavelength, which allows for the use of millimeter and sub-millimeter acoustic resonators implanted deep in tissue \cite{thimot2017bioelectronic}. As shown in Fig.\,\ref{fig:0}, these implants include an integrated circuit (IC) and a piezoelectric crystal (piezo) for power harvesting and communication. Higher power stimulating implants may include an off-chip storage capacitor \cite{piech2020wireless,charthad2018mmsized}. Acoustic power transmitted by an external US transducer is received by the piezo, rectified, and used to power the IC. To send uplink data, the implant can either actively drive its piezo \cite{weber2018miniaturized} or utilize passive backscattering \cite{ghanbari2019sub, sonmezoglu2021monitoring}. In the latter case, modulation of the impedance across the piezo changes the reflection coefficient and backscatter amplitude \cite{ghanbari2020optimizing}. Implant volume is usually dominated by the piezo, which also determines the received power for a given US intensity \cite{ghanbari2020optimizing}. Advanced piezo packaging techniques can improve link efficiency \cite{sonmezoglu2021method}, but the simplest way to reduce volume is to increase the acoustic intensity at the implant. However, this approach is ultimately limited because the FDA restricts the spatial peak-temporal average intensity, $I_{spta}$, of diagnostic US to 720 mW/cm$^2$ \cite{fda2019marketing}.

\begin{figure*}[t]
    \centering
    \includegraphics[width=0.99\linewidth]{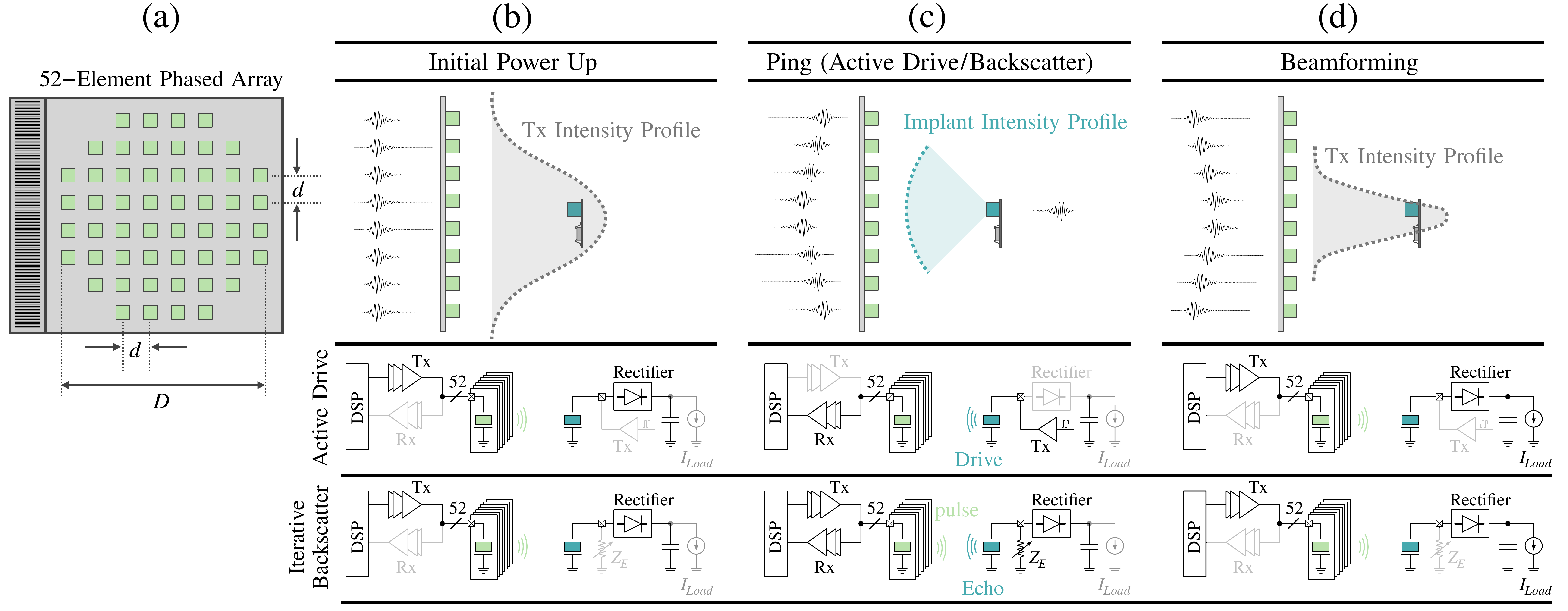}
    \caption[Overview]{(a) Geometry of the 52-element ultrasound array. An 8-element cross section is shown in the rest of the figure. (b) Conceptual diagram of transmitted signals and implant configuration during the unfocused, initial power up. (c) Signals recorded by array and implant configuration during the active drive ping or iterative backscatter. (d) Time-reversed signals transmitted by the array and implant configuration when beamforming.}
    \label{fig:1}
\end{figure*}

%\subsection{Transducer arrays}
Most published US implants use a single-element external transducer. For relatively high power stimulation applications, the greater link efficiency of a focused transducer is preferred, but this setup can tolerate only a few millimeters misalignment \cite{piech2020wireless,charthad2018mmsized}. An unfocused transducer is less sensistive, but it still has a natural focal length, $L_F$, given by its diameter and wavelength. For distances less than $L_F$, in the near-field region, powering an implant is impractical due to local minima and maxima which make received power highly sensitive to alignment and depth \cite{gougheri2019comprehensive}. Thus, implants are typically placed beyond this focal point in the far-field, which provides yet another constraint. While multiple implants in close proximity can be powered using a single transducer, it is preferable to record/stimulate at multiple locations.

To overcome the limitations of a single external transducer, a transducer array (illustrated in Fig.\,\ref{fig:0}) can be used to dynamically focus and steer US by controlling the phase at each element. Linear phased arrays (Nx1 elements) are used in ultrasound imaging to sweep the focus across the azimuth plane, and several have been demonstrated for power delivery to US implants \cite{seo2015ultrasonic,wang2017closed,wang2019ultrasonic,kashani2022design}. However, this still requires manual alignment along one axis. In this work, a planar array (NxM elements) was designed and fabricated since this can steer in both the azimuth and elevation angles to target an implant located within a 3D volume of tissue. Another advantage of a phased array is the ability to power and communicate with multiple implants at different locations; this can be accomplished sequentially through time-division multiplexing \cite{chang2019multi} or simultaneously using techniques such as code-division multiple access \cite{alamouti2020high}. It has been demonstrated that a phased array could be partitioned in half, with each sub-array used to target one implant \cite{wang2017closed, chang2019multi}. This technique was compared to one using the principle of superposition which allows the entire array aperture to be used to simultaneously power multiple implants \cite{benedict2021time}.

Beamforming is often used during signal reception for spatial filtering, and this has been demonstrated in simulation to communicate with a network of implanted sensors in a time-multiplexed fashion \cite{bertrand2014beamforming}. However, this work is focused on transmit beamforming, which can be used to target and efficiently deliver power to implants \cite{seo2015ultrasonic, wang2017closed}. In time delay-and-sum beamforming, the signal transmitted from each array element is delayed based on the distance from each element to the target \cite{hedrick1996beam, wooh1999optimum}. This requires prior knowledge of the implant position relative to the array and the medium's acoustic velocity. To determine the implant location, a subset of array elements can record either a pulse sent by the implant \cite{meng2019self}, or a backscattered signal received from the implant \cite{wang2019ultrasonic}. The time delay for each element can be calculated by finding the maximum of the cross-correlation between the recorded signals. After solving a nonlinear optimization problem to determine the implant location, the delay-and-sum transmit beamforming method can be applied \cite{wang2019ultrasonic}. This approach does not account for tissue inhomogeneity and scattering which may distort and redirect the beam. This work investigates the use of a computationally simple method for beamforming that is inherently robust to tissue inhomogeneity, scattering, and the potentially changing geometry of a flexible transducer array.

Time reversal (TR) beamforming relies on the time reversal invariance of acoustic waves in a lossless medium, and it has been shown to be the optimal solution for maximizing pressure at a target \cite{fink1992time}. Even in some cases with significant attenuation, such as focusing through the skull, a modified TR procedure has been demonstrated \cite{fink2003time}. As a consequence of TR invariance, if an acoustic signal originates from some location and is recorded by an array of transceivers, then playing those recordings backwards from the array elements will result in a reversed version of the original signal converging on the original source. Sending power to an implant using this technique therefore requires a pulse originating from the implant. This can be accomplished using either a single-step “active uplink” method or an “iterative” pulse-echo approach, both illustrated in Fig.\,\ref{fig:1}. The active uplink method requires that the implant can actively drive its piezo. In this case, the implant transmits an acoustic pulse or “ping” which is received by the array (Fig.\,\ref{fig:1}(c)). Recording, reversing, and playing back the signals focuses acoustic power on the implant (Fig.\,\ref{fig:1}(d)). To initially power the implant, the array would start in an unfocused high-power mode (Fig.\,\ref{fig:1}(b)) and/or sweep its focus using standard delay-and-sum beamforming  \cite{benedict2021time}.

However, many low-power ultrasonic implants communicate only through passive backscattering \cite{ghanbari2019sub, charthad2018mmsized,piech2020wireless}, and in this case the iterative method would be used. In this approach, the transducer array first sends out an unfocused pulse (Fig.\,\ref{fig:1}(b)). The implant provides a highly reflective target compared to its surroundings, and the backscatter, or echo, originating from the implant will be received by the array (Fig.\,\ref{fig:1}(c)). This echo is reversed in time and played back (Fig.\,\ref{fig:1}(d)). Iterating over these steps will result in the US beam converging on the strongest reflector \cite{fink2003time}.

One disadvantage of time reversal is that the length of the power pulse from the array is dependent on the length of the ping transmitted from the implant. This is feasible for ultrasonic recording implants because they typically operate with short pulses with a length of up to twice the time-of-flight (ToF) \cite{alamouti2020high}. In the context of delivering power to ultrasonic implants, the acoustic signals will primarily contain a single frequency component, usually set to the resonant frequency of the implant piezo. Therefore, the waveforms can essentially be reduced to their amplitude envelope and phase. However, the ability to arbitrarily vary the pulse length and amplitude envelope provides greater flexibility for communication protocols. Therefore, the use of just phase information is also explored, and this is referred to as phase reversal.

\section{Phased array design}
A 2D, planar transducer array is essential for ensuring implant functionality regardless of lateral misalignment. However, a planar array has greatly increased system complexity compared to a 1D, linear array. As described in Section III, the custom ultrasound system uses MAX14808 high voltage pulsers, controlled using a parallel interface. With the finite number of I/O pins on the FPGA, the number of transducer channels that can be independently driven is limited to 52. This results in an 8x8 element planar array with the three elements in each corner removed, as shown in Fig.\,\ref{fig:1}(a). The limit on channel count motivates the following analysis to determine a suitable choice for array pitch (inter-element spacing).

\subsection{Directivity and grating lobes}
\begin{figure}[t]
    \centering
    \includegraphics[width=1\linewidth]{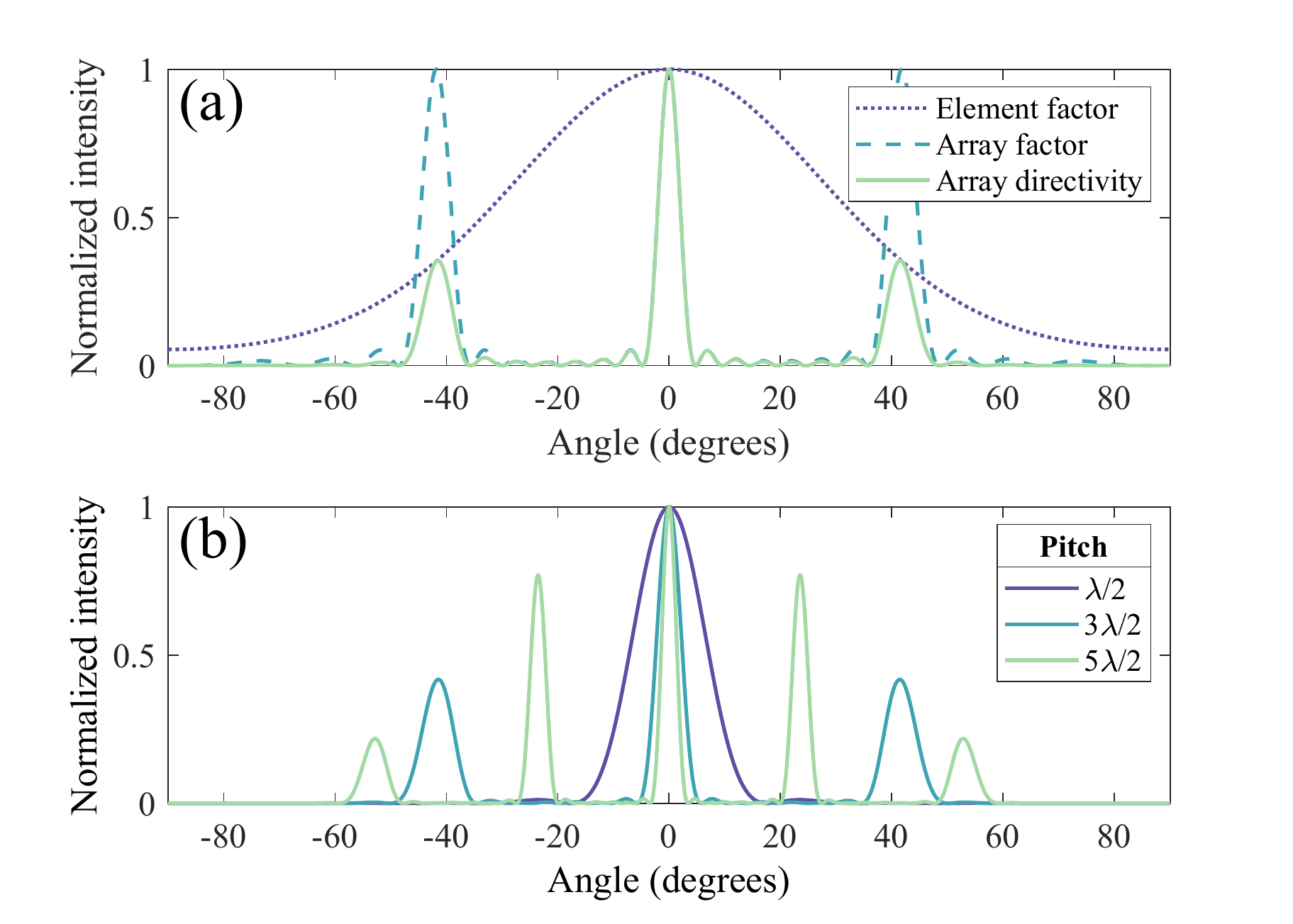}
    \caption[]{(a) Array factor, element factor, and directivity in the far-field for an 8-element linear array with $\lambda$=1 mm, $d$=1.5 mm. (b) Far-field directivity for a 52-element planar array with varying pitch.}
    \label{fig:2}
\end{figure}

The choice of pitch involves a tradeoff between array directivity and the production of grating lobes. The directive gain at a given point is defined as the power density at that point divided by the isotropic power density \cite{mailloux2005phased}. The maximum directive gain is commonly referred to simply as directivity. For a large array with equal element excitation, the far-field radiation pattern/directivity is the product of the array factor (the directivity resulting from the array geometry) and the directivity of the radiation from each element \cite{wooh1999optimum}. The magnitude of the array factor for a linear array with equally spaced point source elements as a function of the angle $\theta$ from the axis is given by:
\begin{equation}
AF = \frac{sin[N\frac{\pi d}{\lambda}(sin\theta - sin\theta_s)]}{sin[\frac{\pi d}{\lambda}(sin\theta - sin\theta_s)]}
\end{equation}
where $\theta_s$ is the desired steering angle, $N$ is the number of elements, and $d$ is the pitch \cite{mailloux2005phased, wooh1999optimum}. This function has maxima for $\theta=\theta_s$, and the array factor value at this angle equals $N$. Thus, the maximum value of the array factor is dependent only on the number of elements. This also holds true for the array factor of planar array, which is approximated by the product of two linear array factors \cite{mailloux2005phased}. An example pattern is shown in Fig.\,\ref{fig:2}(a) for a linear array of 8 elements with a $d$=1.5 mm and $\lambda$=1 mm, with a steering angle of 0$\degree$. The grating lobes are reduced compared to the main lobe due to the element factor. The array factor will be at a maximum when:
\begin{equation}
\theta = sin^{-1}(\frac{m\lambda}{d} + sin\theta_s)
\end{equation}
where $m$ is an integer. Depending on steering angle and ratio of pitch to wavelength, Eq. (2) may have multiple solutions, resulting in grating lobes. Ideally, grating lobes should be minimized to avoid powering an incorrect implant. To eliminate grating lobes at all steering angles, $d<\lambda/2$ should be chosen, but a pitch of up to $\lambda$ can be used without producing grating lobes if steering to $0\degree$.

The beamwidth of a phased array is determined by the total aperture of the array. For a square planar array, the angular beamwidth is $0.866\lambda/D$, where $D$ is the length of a side of the array \cite{mailloux2005phased}. Increasing the array aperture narrows the beamwidth. In Fig.\,\ref{fig:2}(b), acoustic intensity as a function of angle is shown for a 52-element planar array, calculated using the MATLAB Phased Array System Toolbox. While increasing pitch does result in grating lobes and decreased main beamwidth, the maximum intensity, or array gain, is identical for each configuration. Therefore, as long as the beamwidth is sufficiently larger than the implant piezo, the grating lobes produced by a larger pitch array are not detrimental for efficient power delivery to the implant.

\subsection{Optimizing pitch for maximum power delivery}

The previous analyses have considered the far-field radiation pattern of phased arrays; however, the implant could be located in the near-field. The Fresnel region, or near-field, of a circular transducer extends to $\frac{D^2}{4\lambda}$, where $D$ is the transducer diameter \cite{kino1987acoustic}. In this region, a phased array can be used to both steer and focus acoustic power \cite{szabo2013diagnostic}. If an implant is located in the far-field of an array, then the focal point will occur at a shallower depth than the implant. Since the FDA-limited spatial peak-temporal average intensity, $I_{spta}$, will occur at the focal point and not the implant, this ultimately limits the power which the implant can harvest. However, if the implant is located in the near-field, then the acoustic power can be focused at the depth of the implant. The near-field of a 52-element, 1.5 MHz, planar transducer with $d=\lambda/2$ measures only 3 mm, much shallower than an implant. The following analysis will demonstrate why a larger pitch array with grating lobes can be preferable for maximizing power delivered to an implant. Field II is used for simulations \cite{jensen1992calculation, Jensen96field}.

\begin{figure}[t]
    \centering
    \includegraphics[width=1\linewidth]{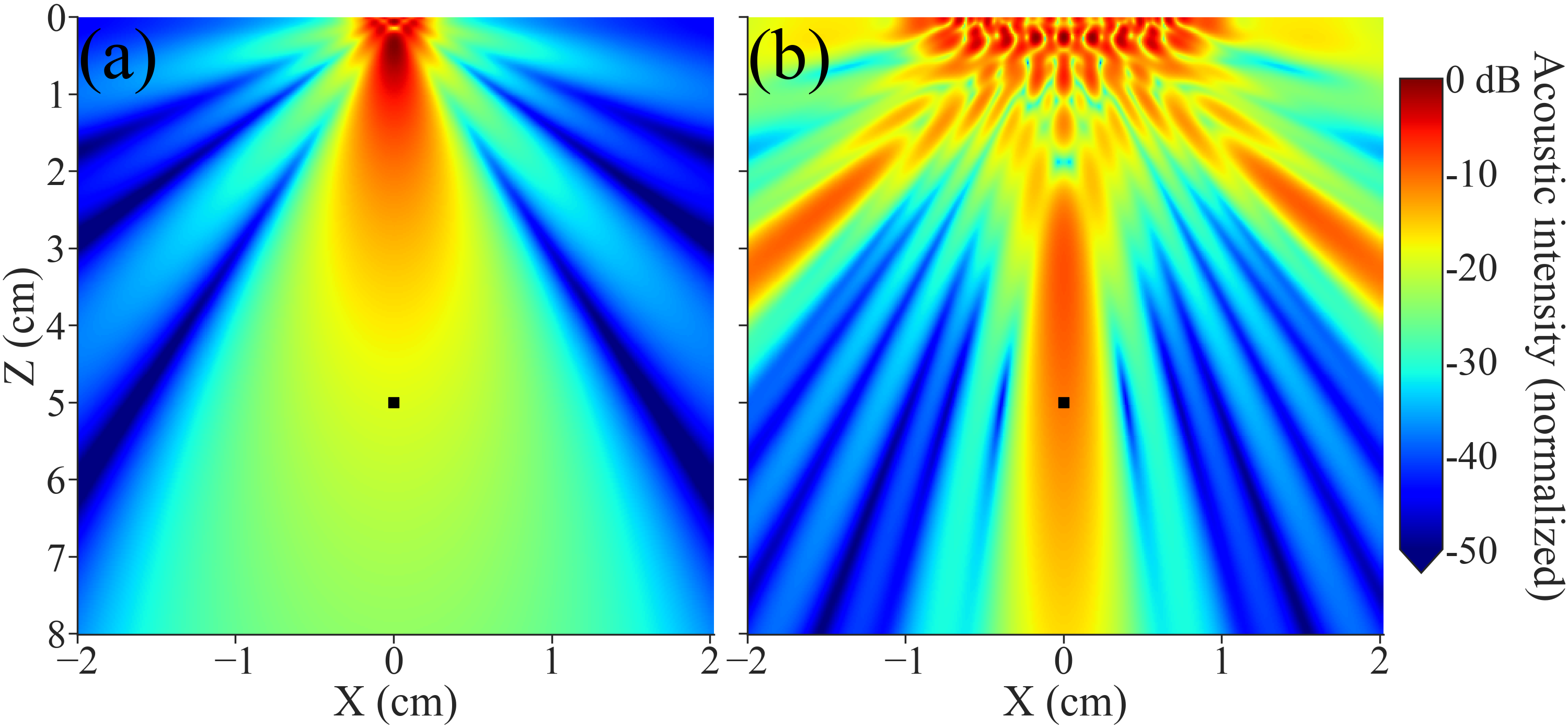}
    \caption[Simulated acoustic intensity when targeting an implant]{Simulated acoustic intensity (in dB) for a 52-element planar array with a pitch of (a) 0.5 mm and (b) 2.0 mm with $\lambda$ = 1.0 mm when targeting a 0.8 mm implant piezo (black square) located at 50 mm depth. Results for each have been scaled to have an equal $I_{spta}$.}
    \label{fig:3}
\end{figure}

The near-field can be extended while maintaining a constant number of elements by increasing the pitch. Acoustic intensity while targeting an implant at 50 mm depth using a 52-element planar array is shown in Fig.\,\ref{fig:3}, with a pitch of $\lambda/2$ = 0.5 mm in (a) and a pitch of $2\lambda$ = 2.0 mm in (b). Both arrays have square elements of size $\lambda/2$ and are targeting an implant located at 50 mm depth (Z axis). The array drive voltage has been scaled such that $I_{spta}$ is equal in both cases. In addition to producing grating lobes, increasing the pitch relative to wavelength leads to a pattern of high intensity regions near the transducer, as seen in Fig.\,\ref{fig:3}(b). Unless an acoustic spacer is used, these high intensity regions must be considered when attempting to maximize power delivered to the implant.

\begin{figure}[t]
    \centering
    \includegraphics[width=1\linewidth]{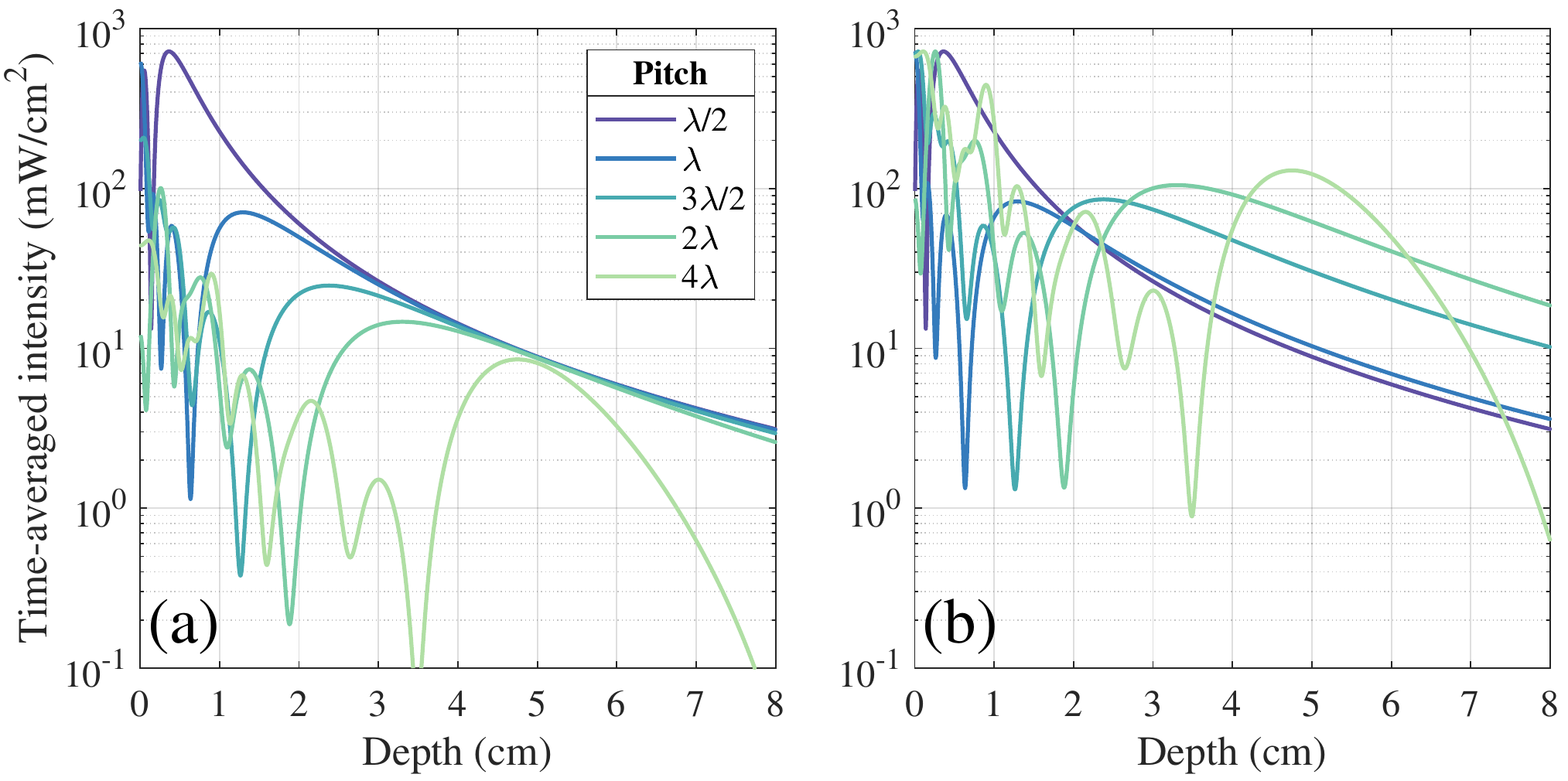}
    \caption[Intensity along axis with constant drive voltage and with scaling]{Simulation of time-averaged intensity along axis while varying pitch ($\lambda$ = 1.0 mm) for a 52-element planar array focused at 50 mm depth. In (a) a constant drive voltage is used for all configurations, while in (b) the drive voltage is scaled for each configuration such that $I_{spta}$=720 mW/cm$^2$.}
    \label{fig:4}
\end{figure}

Fig.\,\ref{fig:4} shows the axial time-averaged acoustic intensity for 52-element planar arrays of increasing pitch focused at a depth of 50 mm. Each uses an element size of $\lambda/2$. A constant element drive voltage is used for each pitch configuration in Fig.\,\ref{fig:4}(a). While the intensity at 50 mm is nearly constant, the peak intensity is lower with a larger pitch. Therefore, for the larger arrays, the drive voltage can be safely increased to increase power delivered to the implant. Fig.\,\ref{fig:4}(b) shows the results of this rescaling which sets the $I_{spta}$ for each configuration equal to the FDA limit of 720 mW/cm$^2$. An array with a 4 mm pitch results in over an order of magnitude improvement in acoustic intensity at the target compared to the array with a 0.5 mm pitch, but this will produce 48 grating lobes when focused at 0$\degree$ based on directivity equations for planar arrays \cite{mailloux2005phased}.

\begin{figure}[t]
    \centering
    \includegraphics[width=\linewidth]{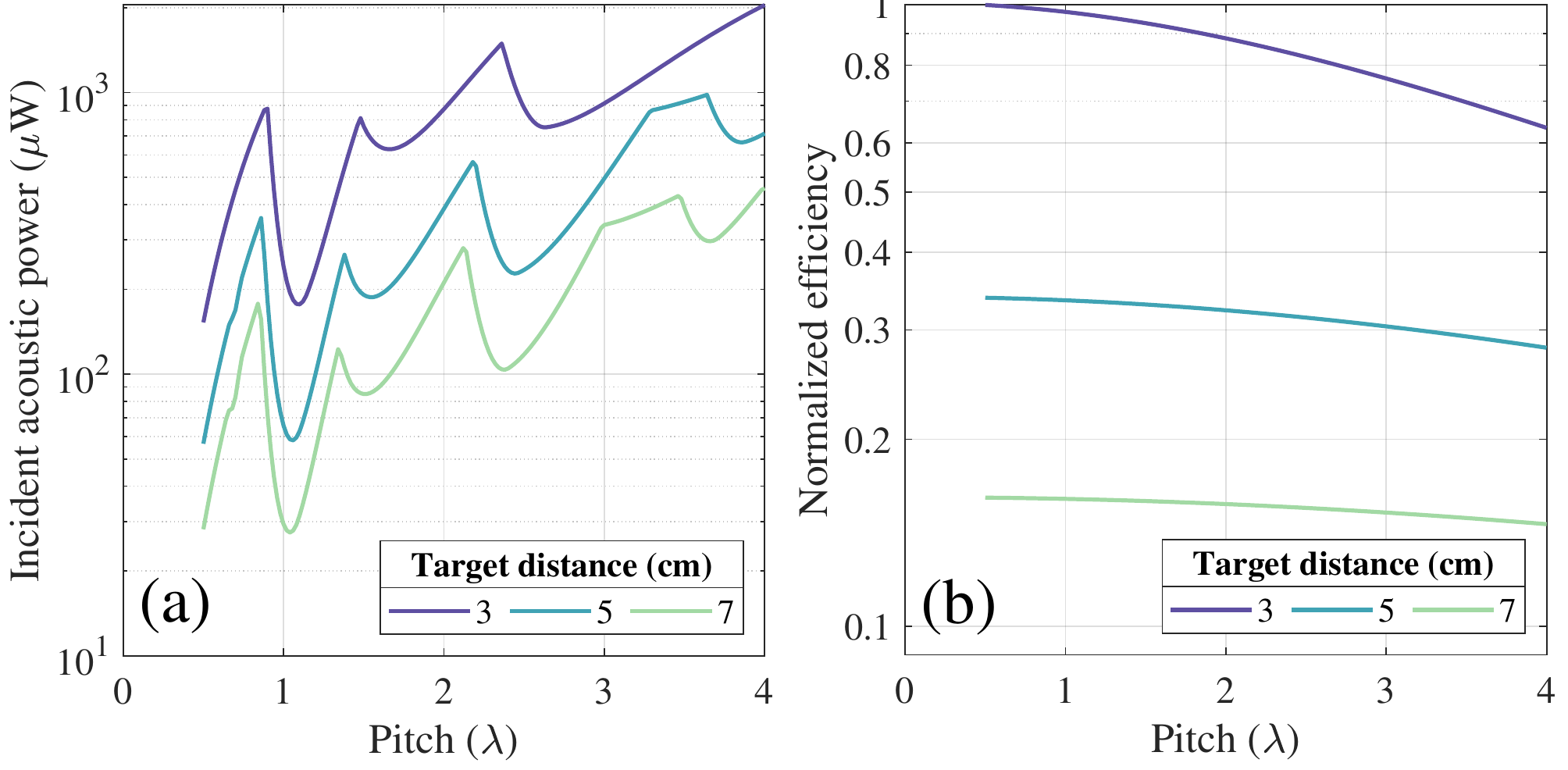}
    \caption[Maximum allowable incident power and efficiency]{(a) Simulated maximum allowable incident acoustic power on a 0.8 mm cube piezo target while meeting the FDA $I_{spta}$ limit. (b) Simulated power transfer efficiency (normalized) from transducer to piezo. ($\lambda$ = 1.0 mm)}
    \label{fig:5}
\end{figure}

The rescaling of drive voltage for larger pitch arrays results in a greater total acoustic power entering the tissue but allows for more power to be delivered to an implant without exceeding the FDA $I_{spta}$ limit. Using this approach to set  $I_{spta}$ to the FDA limit, Fig.\,\ref{fig:5}(a) shows the incident acoustic power on the face of a 0.8 mm cube piezo while varying pitch. This is calculated by integrating the intensity over the piezo face and thus accounts for the decreased beamwidth with increasing pitch. The normalized power transfer efficiency from array to implant is shown in Fig.\,\ref{fig:5}(b). The efficiency does not decrease significantly with increasing pitch. While the results shown are for a steering angle of $0\degree$, they do generalize to larger steering angles, and increasing pitch does not significantly affect steering ability.

The maxima and minima in Fig.\,\ref{fig:5}(a) result from the intensity pattern in the near-field, which is dependent on the exact acoustic properties of the tissue. These results should not be used to determine an optimal pitch value since this will change depending on implant depth and tissue properties. Rather, they lead to the general conclusion that a pitch larger than $\lambda/2$ may be chosen if the goal is to maximize power safely delivered to an implant. This result is particularly relevant for planar arrays, in which channel count is often limited. These results assume ideal focusing of acoustic power on the target. However, even if the array is poorly focused, near-field maxima are present and limit the safe input power. Beamforming efficiency is therefore also important to maximize power at the implant.

\begin{figure}[t]
    \centering
    \includegraphics[width=\linewidth]{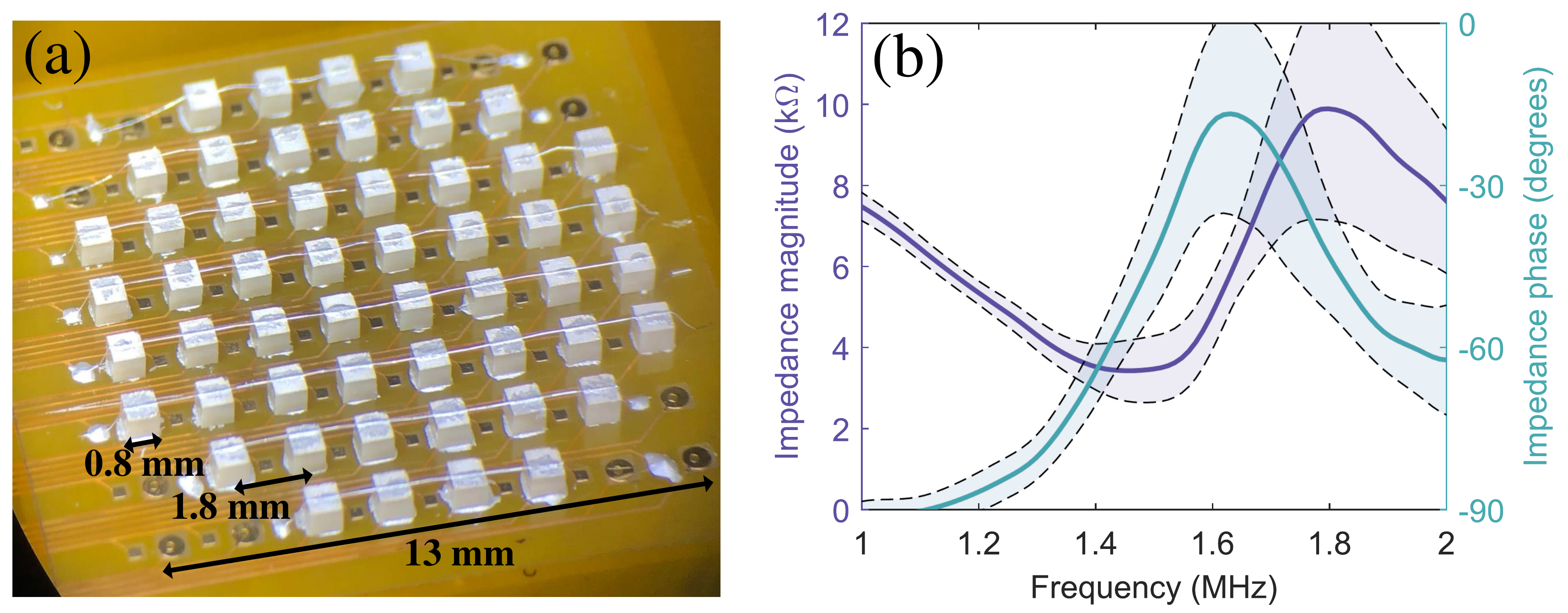}
    \caption[]{(a) Image of fabricated array. (b) Mean and standard deviation of array element impedance measurements in oil.}
    \label{fig:6.5}
\end{figure}

\section{Methods}

\subsection{Array fabrication}
After consideration of the design tradeoffs described in the preceding section, a pitch of 1.8 mm was chosen. With the limited number of elements, this is a good compromise among maximizing link efficiency, maximizing intensity at the implant while meeting FDA limits, and minimizing grating lobes. A total of eight grating lobes will be produced, with four at 33$\degree$ and four at 51$\degree$ from the axis when focused at broadside (0$\degree$) \cite{mailloux2005phased}.

The 52-element, 13 mm diameter planar array was assembled on a 0.3 mm polyimide flexible printed circuit board. Most ultrasound arrays require gel for acoustic coupling, but this is not ideal for long-term applications. A flexible or stretchable array could conform to the skin without the need for gel, as has recently been demonstrated \cite{wang2021continuous}. Lead zirconate titanate (PZT) piezoceramic (APC851) was diced into 0.8 mm cubes, and these elements were attached with silver epoxy (EPO-TEK H20E). The top electrodes were connected using bonding wire and silver epoxy. An image of the array is shown in Fig.\,\ref{fig:6.5}(a). Backing and matching layers could be added to improve efficiency and protect the elements \cite{gougheri2019comprehensive}. Impedance measurements revealed 4 defective elements; the remainder showed good matching with each other and with finite element model simulations \cite{ghanbari2020optimizing}. The impedance of a typical element is shown in Fig.\,\ref{fig:6.5}(b). The series resonant frequency of these piezos was measured as 1.48 $\pm$ 0.08 MHz; the array was operated at 1.5 MHz.

\subsection{Ultrasound system design}
\begin{figure}[t]
    \centering
    \includegraphics[width=\linewidth]{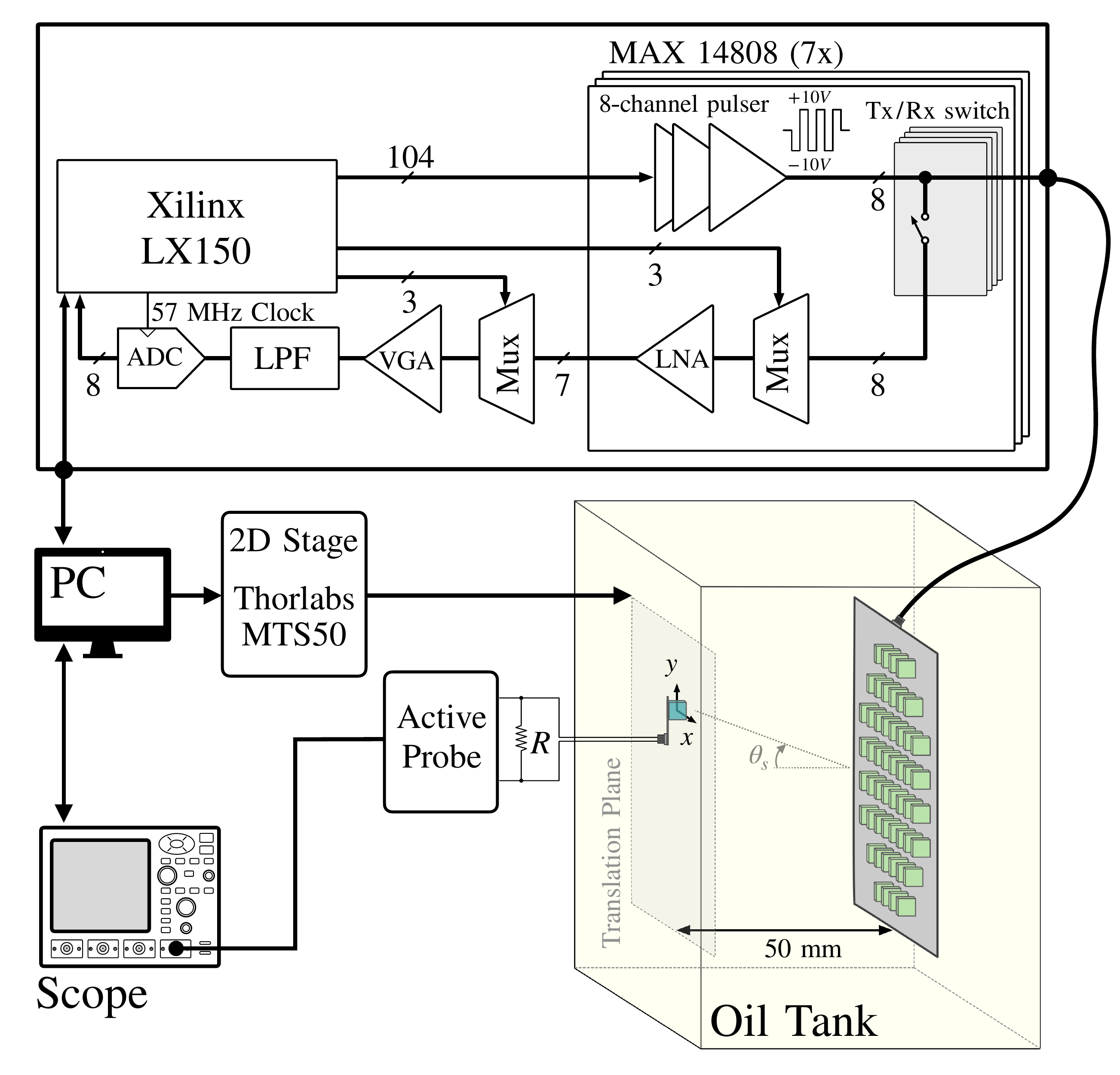}
    \caption[]{Custom ultrasound beamforming system and experimental setup.}
    \label{fig:6}
\end{figure}

The custom US system (diagrammed in Fig.\,\ref{fig:6}) used the MAX14808, a three-level high-voltage digital pulser with integrated transmit/receive switches. The pulser supply was automatically varied from $\pm$5 to $\pm$25 V based on the amplitude of backscattered signals. The variable gain amplifier (VGA) gain was constant for all channels and was calibrated once at startup to ensure no channel was saturating while recording reflected signals from the implant. All energy transfer efficiency measurements were taken using a $\pm$10 V supply ($\pm$8.3 V driving piezos), with $\sim$8.6 mW consumed by each element. A Spartan-6 LX150 FPGA was used to generate the pulser control signals, control the receive path multiplexers, acquire data at 57 MHz from the ADG9047 8-bit analog to digital converter (ADC), and communicate with a PC over USB. OPA355 op amps were used for the low-noise amplifier, programmable gain amplifier, and anti-aliasing filter. Since only one ADC was used, recording from the entire array required recording 52 identical pulses. This is acceptable for operation in iterative mode and for this demonstration of active uplink reversal; however, to use active uplink reversal with a true implant requires at least one additional ADC.

\subsection{Experimental setup}
The model implant was a 0.8 mm cube PZT piezo mounted on a small flex-PCB. The perpendicular distance between the array and implant, or implant depth, was 50 mm. Motorized translation stages (Thorlabs MTS50) controlled the lateral position of the implant. This setup is diagrammed in Fig.\,\ref{fig:6}. The array and “implant” piezo were submerged in canola oil (c $\approx 1470$ m/s, $\rho\approx 910$ kg/cm$^3$, $\alpha\approx0.15$ dB/cm), which has reasonably similar acoustic properties to tissue. For experiments that characterized performance through tissue, $\sim$25 mm porcine tissue (c $\approx 1580$ m/s, $\rho\approx 1070$ kg/cm$^3$, $\alpha\approx 2$ dB/cm) was suspended in the oil between array and implant.

The implant piezo was wired to a multiplexer to control the piezo load impedance. During energy transfer efficiency measurements, a 2.5 k$\Omega$ resistor was connected to the piezo terminals; this matched load allows for maximum power transfer at the series resonant frequency. The voltage across the resistor was recorded using an oscilloscope and active differential probe (Keysight N2750A), and this was used to calculate the energy delivered to the load. The total energy transmitted from the array was estimated from the voltage waveforms applied to the 52 array elements and the element impedance at resonance. Energy transfer efficiency was found by dividing received energy by transmitted energy. Efficiency was used to compare methods because the applied voltage waveforms for time reversal were not explicitly controlled and could slightly differ in duration and input energy.

\begin{figure}[t]
    \centering
    \includegraphics[width=\linewidth]{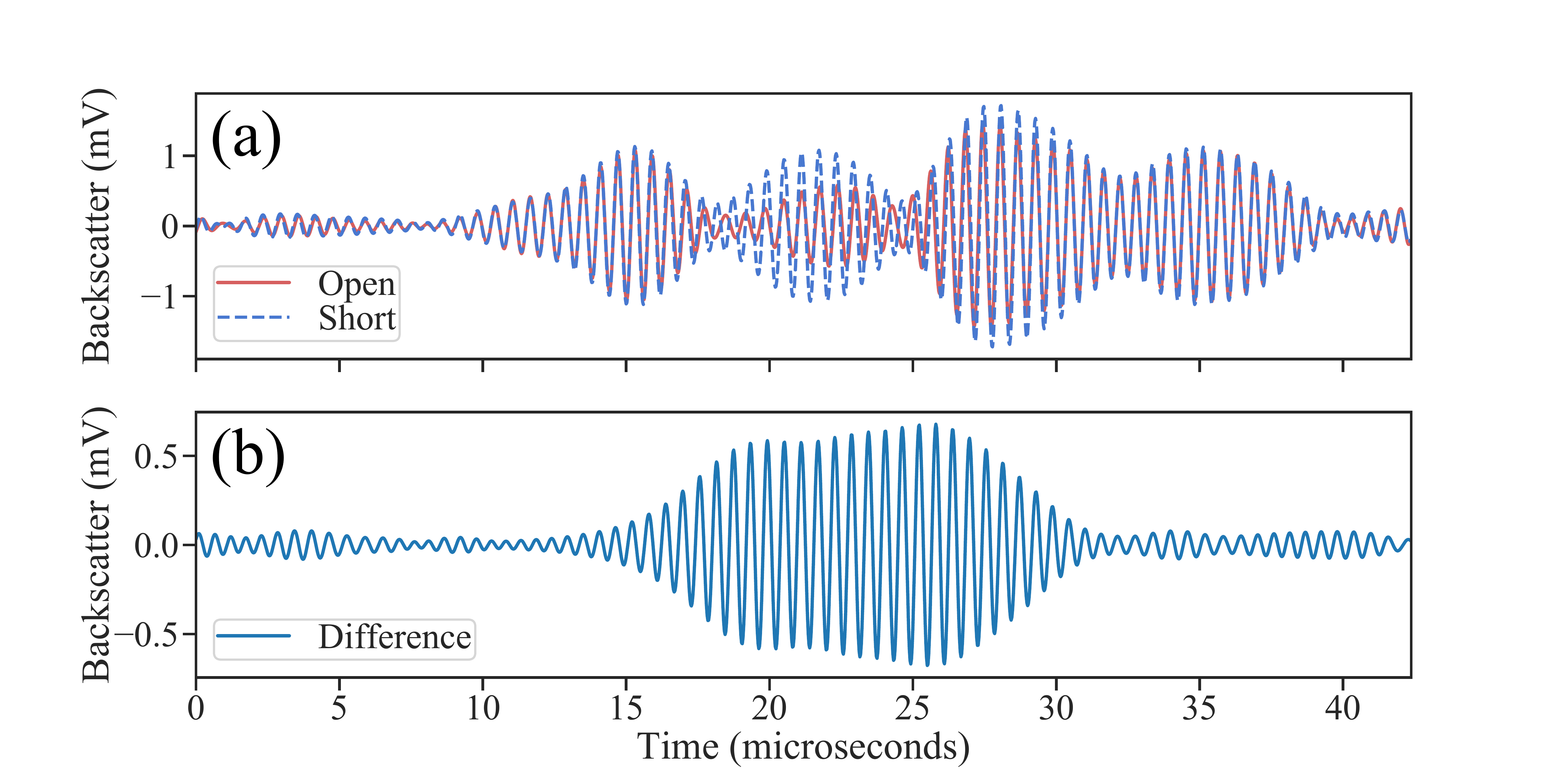}
    %\caption[]{(a) Received backscatter on an array element with implant piezo open and shorted. (b) Difference between open and shorted signals}
    \caption[]{(a) Received backscatter on an array element with implant piezo open and shorted. Undesirable reflections are present. (b) Difference between open and shorted backscatter, consisting primarily of the desired implant reflection.}
    \label{fig:7}
\end{figure}

\subsection{Delay-and-sum beamforming}
For delay-and-sum beamforming, required element delays were calculated from the known location of the implant relative to the array and the acoustic velocity. The implant was first aligned with the array by focusing at broadside and moving the implant to the position with maximum received power. This location was then used as the (0,0) position in the X and Y axes for all experiments.

\subsection{Active uplink reversal}
For active uplink reversal, the implant piezo was driven with $\pm$3.3 V for 40 cycles. The pulse was received by the array and bandpass filtered. The beginning and end of the signals were identified by a threshold crossing, and the signals were rescaled to compensate for variation in piezo response among elements. No assumptions were made about implant depth. For time reversal, the waveforms were reversed and quantized for the 3-level (-$V_{High}$, 0, +$V_{High}$) pulsers. This entire process could be implemented with digital processing and memory on-chip.

To calculate the relative delays between elements for phase reversal, an element at the array center was used as a reference. The maximum of the cross-correlation was found between this reference signal and the signals on each element to determine relative delays. This sometimes resulted in $2\pi$ offsets between elements, or “phase gaps" \cite{montaldo2004revisiting}. For a long pulse, these $2\pi$ offsets would be insignificant, but for shorter pulses it is beneficial if the beginnings of the signals from all elements arrive at the same time. Therefore, element delays were shifted by increments of $2\pi$ if they differed significantly from their neighbors. Use of nearest-neighbor cross correlation could reduce the need for this cleanup procedure \cite{flax1988phase}. Once element delays were found, they were reversed and used to generate the transmit waveforms.

\subsection{Iterative reversal}

The iterative TR process requires that the strongest echo originates from the target, which in this case is the front of the piezo. However, other reflectors in tissue such as bone can create stronger reflections. The piezo is also not the only reflector on the implant, since it also includes an IC, capacitor, and flex-PCB substrate. Preliminary results showed that defocusing could occur as a result of reflections from the flex-PCB, and it was also difficult to identify the low-amplitude backscatter when the implant was at larger angles from the array. To achieve contrast with the surroundings during implant localization, \cite{wang2019ultrasonic} utilized the 3rd harmonic produced by the rectifier connected to the piezo while \cite{zhang2019powering} opened and shorted the piezo while sweeping the array focus during US B-mode imaging. Here, a related approach to the latter was used to identify the reflected signal originating solely from the piezo itself rather than its packaging or surroundings.

The backscatter received by a single array element after sending an unfocused pulse is shown in Fig.\,\ref{fig:7}(a), both with the implant (located $20 \degree$ from broadside) piezo terminals open and shorted. The received echos were composed of multiple reflections, and it was not possible to isolate the piezo reflection. However, calculating the difference between the waveforms, i.e. the backscatter modulation between the two states, results in the signal in Fig.\,\ref{fig:7}(b) which is known to originate from the piezo. Using this difference waveform for TR allows for reliable focusing. To modulate piezo termination, a real implant would still need to be powered on from the unfocused US (Fig.\,\ref{fig:1}(b)); however, this can be done without actively driving the piezo and with power-intensive blocks shut down.

\begin{table}[t]
\centering
%\vspace{-0.5em}
\footnotesize
\renewcommand{\arraystretch}{0.9}
\setlength{\tabcolsep}{3pt}
\caption{Description of beamforming methods}
\begin{tabular}{@{}ll@{}}\hline
\toprule
\bf Method &\bf Description \\
\midrule
Time reversal & Received signals from "ping" are reversed in time\\
(active uplink) & and transmitted by array. \\
\midrule
Phase reversal  & Delays are calculated from recorded signals from "ping." \\
(active uplink) & Array transmits signals with the reverse of those delays. \\
\midrule
Time reversal &  Delays are calculated from recorded backscatter signals. \\ (iterative) & Array transmits signals with the reverse of those delays, \\
& multiplied by the recorded amplitude envelope.\\
\midrule
Phase reversal & Delays are calculated from recorded backscatter signals.\\ (iterative) & Array transmits signals with the reverse of those delays. \\
\midrule
Delay-and-sum & Using the known implant position, delays are calculated \\
 & from geometry.\\ 
 \midrule
Unfocused & Array elements all driven with the same signal.\\
\bottomrule
\end{tabular}
%\vspace{-2em}
\label{table:0}
\end{table}

As with active uplink reversal, the waveforms were bandpass filtered, and the beginning and end of the backscattered signal were identified by a threshold crossing. Ten reversal iterations were performed; this was more than sufficient for convergence on the target implant. However, directly performing time reversal was not feasible using the iterative method because the frequency of the signal tended to shift slightly after multiple iterations; this is believed to have been caused by mismatch in the resonant frequency between array elements. Since the acoustic pressure to voltage transfer function of the piezo is dependent on frequency, any shift could obfuscate the comparison between beamforming methods. Therefore, iterative TR was performed by taking the signals used for phase reversal and multiplying them by the amplitude envelope of the received echos. This uses both the phase and amplitude information included in these signals, and a similar reconstruction technique has been used previously \cite{montaldo2004revisiting}. A summary of the process for each beamforming method described is given in Table\,\ref{table:0}.

Iteratively focusing on multiple implants can also be accomplished using the calculated backscatter modulation signal. By sending a downlink command to one implant at a time to open and short its piezo, focusing on each implant can be achieved.

\subsection{Finite-element model simulations}

For comparison to Field II simulation results of array directivity and for estimation of link efficiency, several finite-element model (FEM) simulations were performed in COMSOL Multiphysics. The implant piezo used to experimentally measure the directivity of the array has its own directivity function which must be accounted for when comparing to Field II simulations of the array. The power received as a function of acoustic field incident angle was simulated using a 2D FEM of a 0.8 mm PZT piezo in an acoustic medium. The Field II directivity results were then multiplied by this function to match the experimental setup.

To estimate the electric to acoustic conversion efficiency of the array elements ($\eta_{TX}$), the simulated acoustic power transmitted from the front face of a piezo was divided by the total electrical power consumed. To estimate the acoustic to electric conversion efficiency of the implant ($\eta_{RX}$), the simulated electrical power delivered to a matched load was divided by the incident acoustic power on the piezo face. These two FEM simulations used a 2D axisymmetric model of a 0.8 mm piezo.

\section{Results}

\subsection{Acoustic field characterization}
\begin{figure}[t!]
    \centering
    \includegraphics[width=\linewidth]{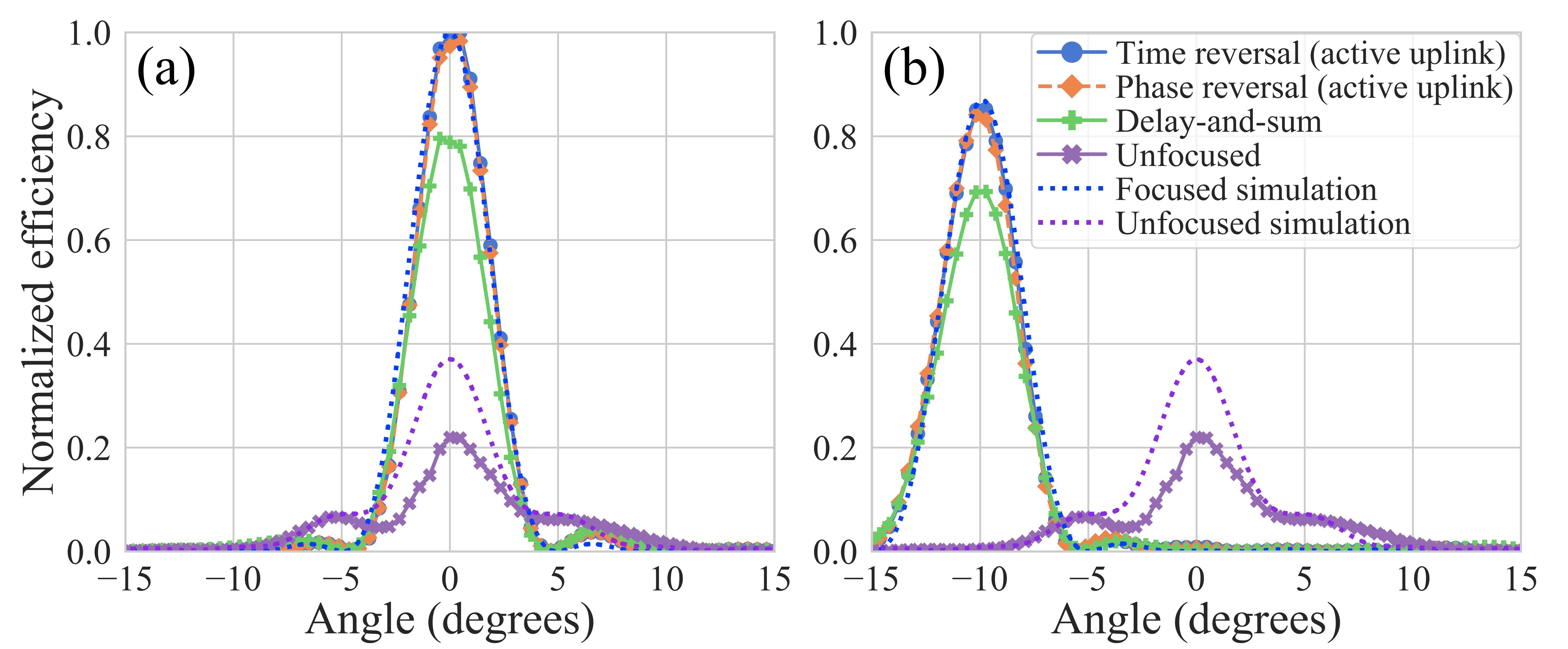}
    \caption[]{Measured and simulated efficiency (normalized to peak) as a function of angle when focused to (a) 0$\degree$ and (b) -10$\degree$ at 50 mm depth. Both plots are normalized to the peak efficiency of 0.024\%}
    \label{fig:8}
\end{figure}

The directivity pattern of the fabricated array was measured and compared to Field II simulation results, as shown in Fig.\,\ref{fig:8}. Since the orientation of the piezo receiver relative to the array was fixed, one of the factors leading to decreased efficiency at larger steering angles was increasing angular misalignment. Therefore, the Field II simulation results of the array directivity have been multiplied by the implant directivity from FEM simulations to match the experimental setup. In Fig.\,\ref{fig:8}(a), acoustic power was steered to 0\degree, while in Fig.\,\ref{fig:8}(b) the beam was steered to -10\degree. Excellent matching was observed between the focused simulation and measurements. The measured half-power beamwidth was 3.9\degree (3.3 mm diameter at 50 mm depth), which was consistent with the 3.7$\degree$ theoretical beamwidth for this array \cite{mailloux2005phased}.

\subsection{Link efficiency and available power}
The power transfer link efficiency primarily consists of the following factors: electric to acoustic conversion efficiency of the array ($\eta_{TX}$), focusing efficiency ($\eta_{foc}$), attenuation of the acoustic medium ($\eta_{att}$), and acoustic to electric conversion efficiency of the implant ($\eta_{RX}$). The total efficiency ($\eta_{link}$) is the product of these factors and was measured to be -36 dB (0.024\%) with the implant piezo centered at 50 mm depth.

From FEM simulations, $\eta_{TX}$ and $\eta_{RX}$ were estimated to be -6 dB and -3 dB, respectively. The focusing efficiency ($\eta_{foc}$) was calculated by dividing the incident acoustic power on the face of the piezo by the integrated acoustic power (including in grating lobes) when sweeping the receiver over the entire 2D plane. This was done both for Field II simulations and experimental measurements, and the results were consistent at -25 dB. Attenuation was low in the oil medium; at 50 mm depth $\eta_{att} \approx$ -1 dB. The final -1 dB may be due to angular misalignment, resonance mismatch, or defective elements.

The measured efficiency corresponds to 110 $\mu$W power delivered to the load. If the array drive power was increased to meet the FDA $I_{spta}$ limit, 225 $\mu$W available power is expected from Field II simulations and the simulated $\eta_{RX}$. This is sufficient for a broad range of ultrasonically-powered implanted physiological sensors and neurostimulators \cite{ghanbari2019sub,sonmezoglu2021monitoring,shi2020monolithically,piech2020wireless}.

\subsection{Comparison of beamforming methods}
\begin{figure*}[t]
    \centering
    \includegraphics[width=\linewidth]{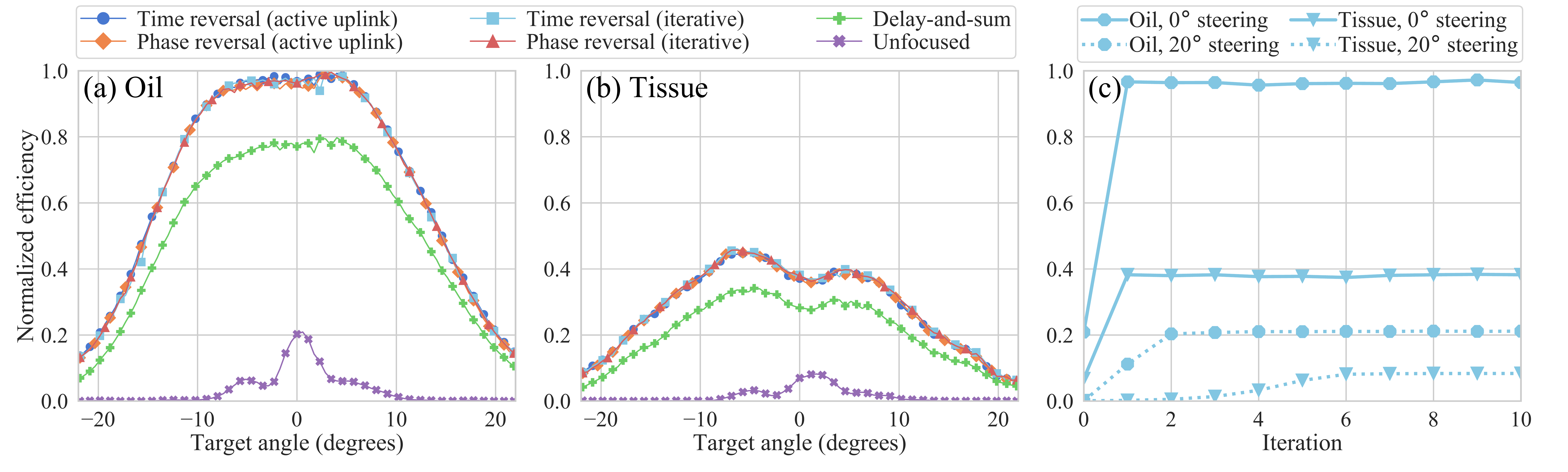}
    \caption[]{Efficiency versus beamforming angle (normalized to peak value) when focusing at each position using each beamforming method through (a) oil medium and (b) porcine tissue suspended in the oil medium. (c) Efficiency after each time reversal iteration through oil and tissue demonstrating the required iterations for convergence. All plots are normalized to the peak efficiency of 0.024\%}
    \label{fig:9}
\end{figure*}

Iterative time and phase reversal are compared to active uplink time and phase reversal in Fig.\,\ref{fig:9}(a) and (b). Results with delay-and-sum beamforming using the known implant location and without beamforming are also shown. Efficiency decreased at larger steering angles due to increased angular misalignment and additional attenuation from greater propagation distance. Measurements were taken at 50 mm depth while sweeping the implant piezo along one axis of the array. Results through the oil medium are shown in Fig.\,\ref{fig:9}(a), while Fig.\,\ref{fig:9}(b) shows results through porcine tissue. The implant was again at 50 mm depth, and results are normalized to the same peak value as in Fig.\,\ref{fig:9}(a). Efficiency was lower through tissue due to greater attenuation in tissue and reflections at the oil-tissue interfaces.

\begin{figure}[t]
    \centering
    \includegraphics[width=\linewidth]{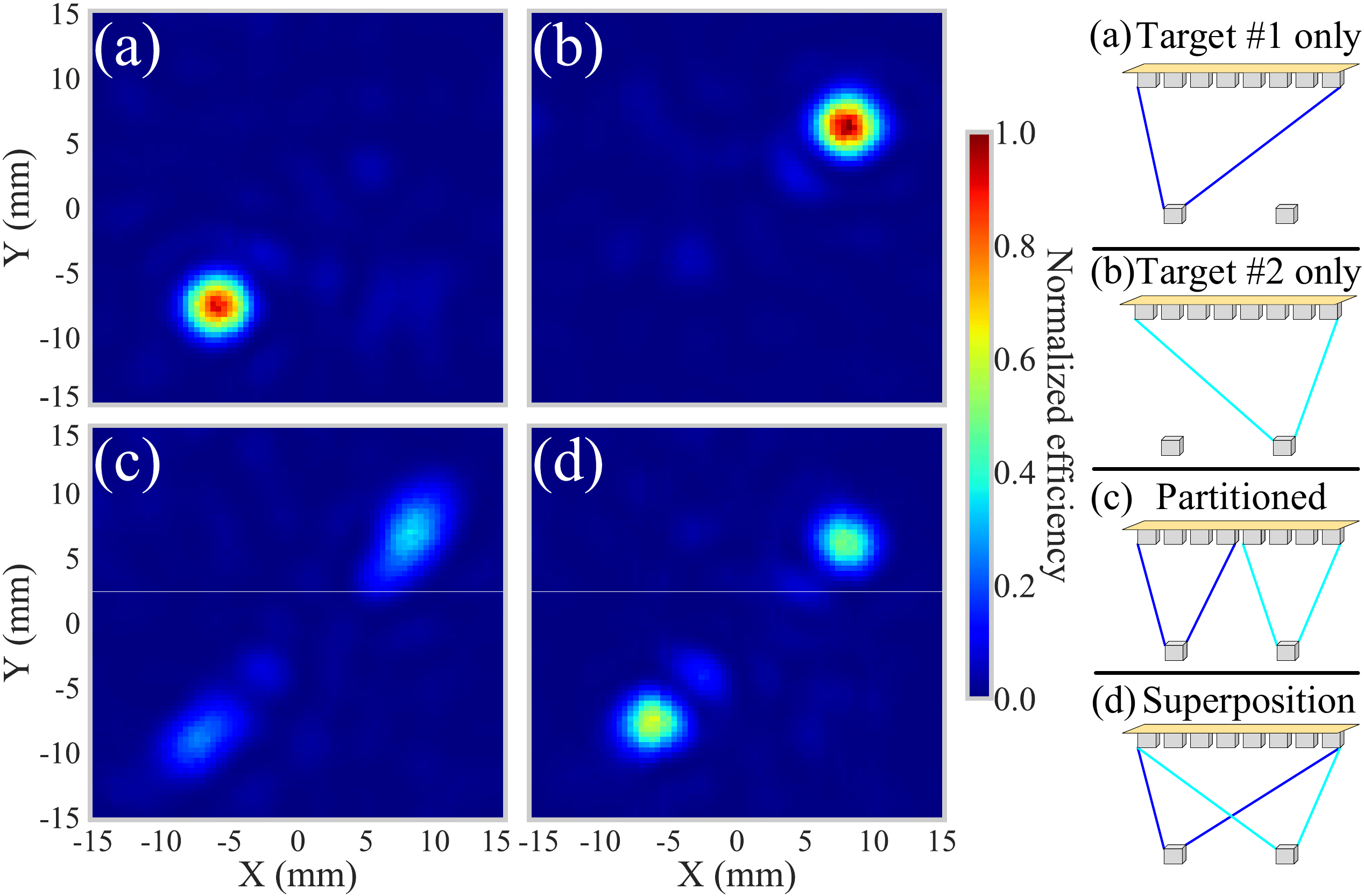}
    \caption[]{Energy transfer efficiency (normalized) with two implants at 50 mm depth with separate and simultaneous time reversal. Illustrations demonstrating each setup are also shown. (a) Target \#1 (-6 mm, -8 mm) only. (b) Target \#2 (8 mm, 6 mm) only. (c) Partitioned array. (d) Superposition. Results are normalized to 0.015\% efficiency. Reproduced from \cite{benedict2021time}.}
    \label{fig:10}
\end{figure}

Time reversal had less than 1\% greater efficiency than phase reversal. Iterative time/phase reversal were able to achieve nearly identical efficiency to active uplink time/phase reversal. All four of these methods were around 20\% more efficient than delay-and-sum beamforming. This difference can likely be attributed to the fact that delay-and-sum beamforming did not account for imperfections in array geometry and fabrication. While this advantage may not be present if using a conventional US imaging array, this self-correction would be useful if the phased array was designed as part of a flexible, wearable device. All methods improved efficiency compared to the unfocused array by 1-2 orders of magnitude.

The number of iterations required for iterative reversal depended on the backscatter strength, as shown in Fig.\,\ref{fig:9}(c). If the initial echo was strong, as was the case when the implant was centered, only one iteration was required. When the implant was off-center, more iterations were required. Still, convergence was quickly achieved, which demonstrates the feasibility of tracking a moving implant in real-time.

\subsection{Multiple implants}
Two implants at 50 mm depth were powered separately (Fig.\,\ref{fig:10}(a),(b)), together using a partitioned array with half focusing on each implant (Fig.\,\ref{fig:10}(c)), and together using superposition (Fig.\,\ref{fig:10}(d)). Active uplink TR was used to generate the transmitted waveforms. Superposition resulted in 60\% and 43\% efficiency at the targeted implants compared to powering each implant separately. This was expected since the acoustic energy was split between two foci and a perfect superposition was not possible due to pulser quantization. The partitioned array suffered from increased beamwidth due to the reduced aperture of each sub-array, resulting in 33\% and 21\% efficiency at each implant.

\begin{figure}[t]
    \centering
    \includegraphics[width=\linewidth]{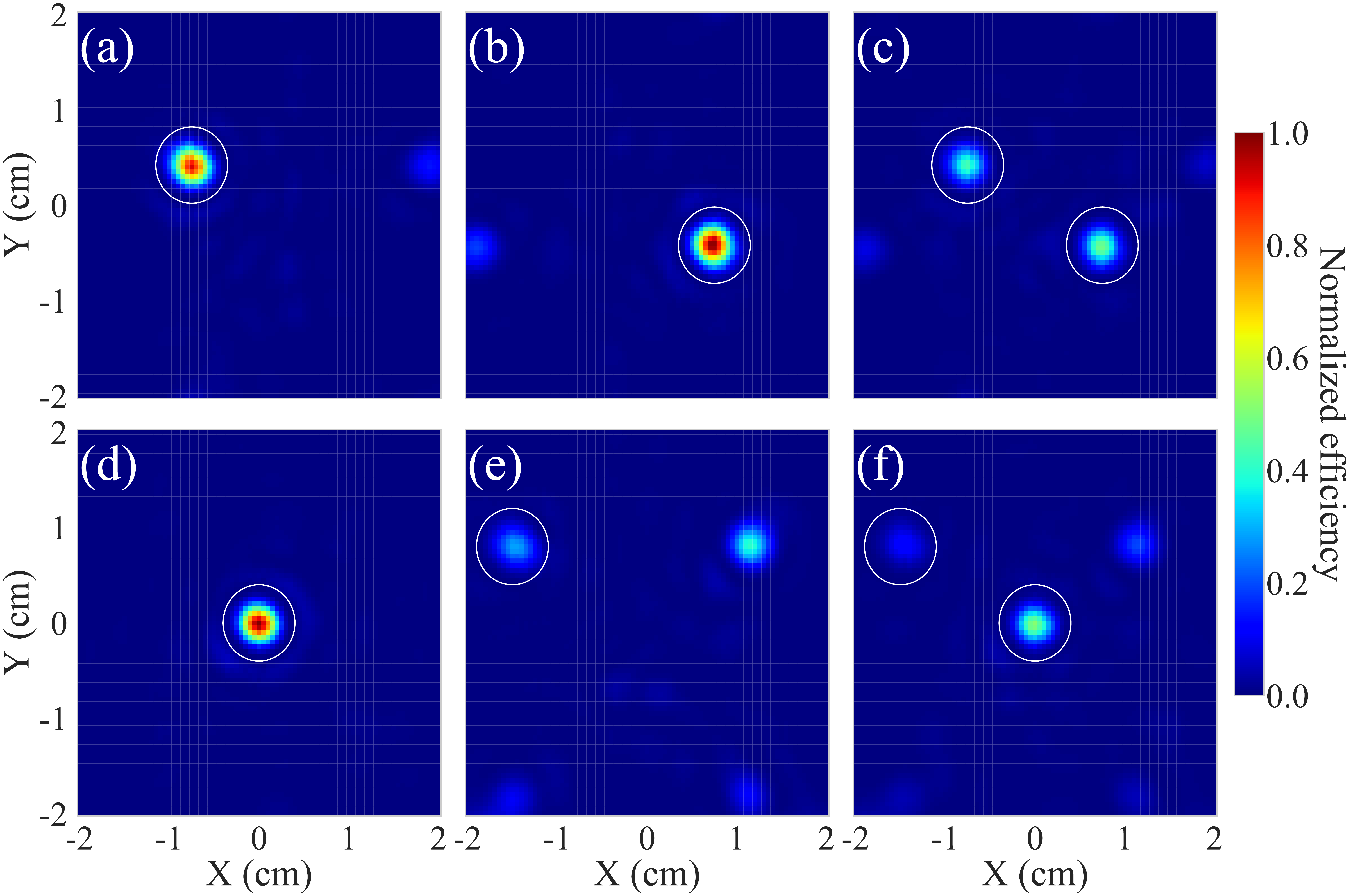}
    \caption[]{Energy transfer efficiency to multiple targets after iterative phase reversal is shown. Main lobes are indicated by white circles, and other lobes are grating lobes. (a) and (b) show phase reversal to separate targets at symmetrical locations. (c) shows superposition to these symmetrical locations. (d) and (e) show phase reversal to separate targets in asymmetrical locations. (f) shows superposition to these asymmetrical locations. Results are normalized to the peak efficiency of 0.024\%}
    \label{fig:11}
\end{figure}

Simultaneous power delivery to two implants using superposition after iterative reversal is shown in Fig.\,\ref{fig:11}. The required delays to target each implant were determined using the calculated backscatter modulation signal. The two sets of signals were then summed to simultaneously target both implants. The main lobes are indicated by white circles, and the other regions receiving power are grating lobes. If a grating lobe was closer to the array center than the main lobe, it was possible for a grating lobe to have the greater intensity, as was the case in Fig.\,\ref{fig:11}(e). With symmetrically located implants in Fig.\,\ref{fig:11}(c), the efficiency to each implant was 47\% and 48\% compared to powering each implant separately as in Fig.\,\ref{fig:11}(a)-(b). When one implant was centered and one was off-center, as in Fig.\,\ref{fig:11}(f), the efficiency was 51\% and 47\%, respectively, compared to powering them separately in Fig.\,\ref{fig:11}(d)-(e).

\section{Conclusion}
In this work, a planar phased array was used for ultrasound power delivery to biomedical implants, greatly reducing the sensitivity of the system to misalignment compared to a single-element transducer. The practical limit on channel count and the FDA-mandated diagnostic ultrasound intensity limit motivated the design of an array with a pitch greater than $\lambda/2$. This was shown through simulation to lead to only a small decrease in link efficiency while greatly increasing the maximum power which could safely be delivered to an implant. Simulation results showed that for arrays with a large pitch, the intensity peaks that limit the array input power occur in the near-field diffraction pattern rather than at the focus. Therefore, improving beamforming efficiency directly translates to increased power available to the implant. Increased efficiency is also beneficial due to the power-limited environment of a wearable device and the desire to dissipate the minimum amount of acoustic energy in tissue while powering chronic implants.

Time reversal provides a theoretically optimal approach to beamforming and accounts for tissue inhomogeneity and the changing geometry of a flexible or stretchable array. Using a custom array, TR beamforming was demonstrated and compared to other beamforming techniques. This work showed that time reversal can be used both in an active uplink mode if the implant communicates by actively driving its piezo and in an iterative pulse-echo mode if the implant communicates through backscattering. Time reversal and the related phase reversal approach both showed approximately 20\% better energy transfer efficiency than delay-and-sum beamforming using the known implant location. Superposition was used to simultaneously power two ultrasonic implants, and this was twice as efficient as using half the array to power each implant. Finally, this work demonstrated that time reversal can be used with reflective implants by using the backscatter modulation signal as a method of locating and powering multiple implants with iterative time reversal.

While this work demonstrated TR beamforming using a custom planar phased array, this method can be used with any 1D or 2D ultrasound array. However, planar arrays are advantageous for ambulatory, wearable applications to correct for implant alignment and migration in 3D. Using a planar array, however, results in large channel counts that increase the size, complexity, and power dissipation of the system. The eventual development of a compact phased array system for powering a network of miniaturized implants would therefore require high efficiency power delivery and simple on-chip processing enabled by the strategies proposed in this work.

\appendices
\section*{Acknowledgment}
The authors thank Sina Faraji Alamouti, Nathan Tessema Ersumo, Dr. Soner Sonmezoglu, Professor Steven Conolly, and Professor Chris Diederich for technical discussions.

\ifCLASSOPTIONcaptionsoff
  \newpage
\fi


\bibliographystyle{IEEEtran} %
\bibliography{ref-extracts} % Entries are in the "refs.bib" file

\vspace{-10cm}
\begin{IEEEbiography}[{\includegraphics[width=1in,height=1.25in,clip,keepaspectratio]{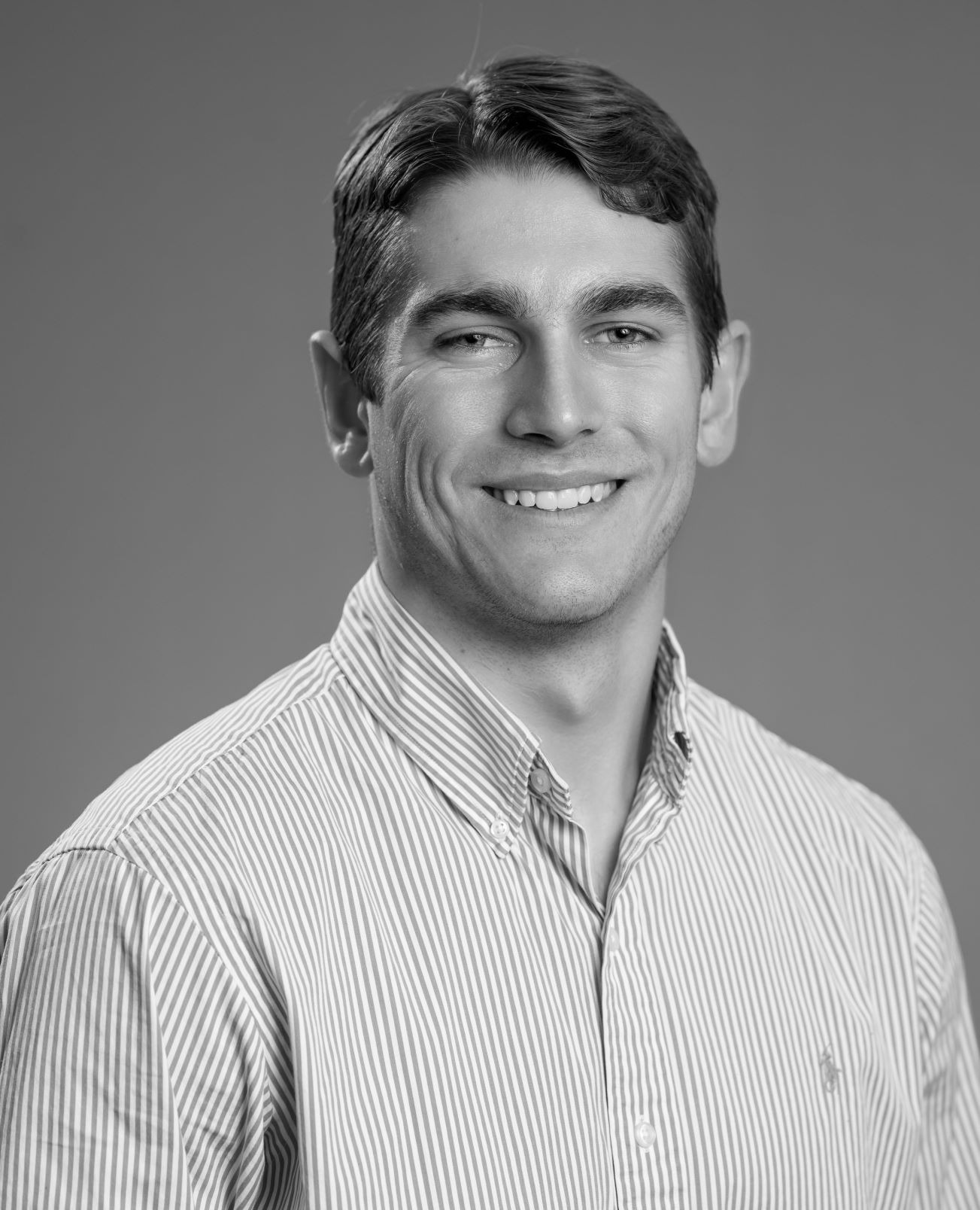}}]{Braeden C. Benedict} (Student Member, IEEE) received the B.S. degree in electrical engineering from the University of Notre Dame in 2019. He is a recipient of the National Science Foundation Graduate Research Fellowship. In 2021, he received the M.S. degree in bioengineering from the University of California, Berkeley–University of California, San Francisco Joint Graduate Program in Bioengineering. His research interests include implantable medical devices and neural interfaces.
\end{IEEEbiography}

\newpage

\begin{IEEEbiography}[{\includegraphics[width=1in,height=1.25in,clip,keepaspectratio]{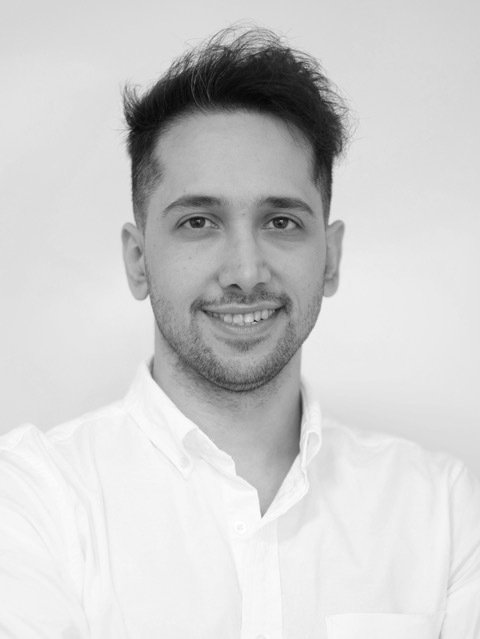}}]{Mohammad Meraj Ghanbari} (Student Member, IEEE) received the M.Eng. and M.Phil. degrees, both in Electrical Engineering, in 2013 and 2016 from the University of Melbourne, Victoria, Australia. He is currently a Ph.D. candidate at the Electrical Engineering and Computer Science department at the University of California, Berkeley, CA, USA. His research interests include low power analog and mixed-signal integrated circuits, data converters, energy harvesting, biosensing and neural interfaces. He is a recipient of the ADI Outstanding Student Designer Award in 2019 and the Apple Ph.D. Fellowship Award in 2021.
\end{IEEEbiography}

\enlargethispage{-10cm}

\begin{IEEEbiography}[{\includegraphics[width=1in,height=1.25in,clip,keepaspectratio]{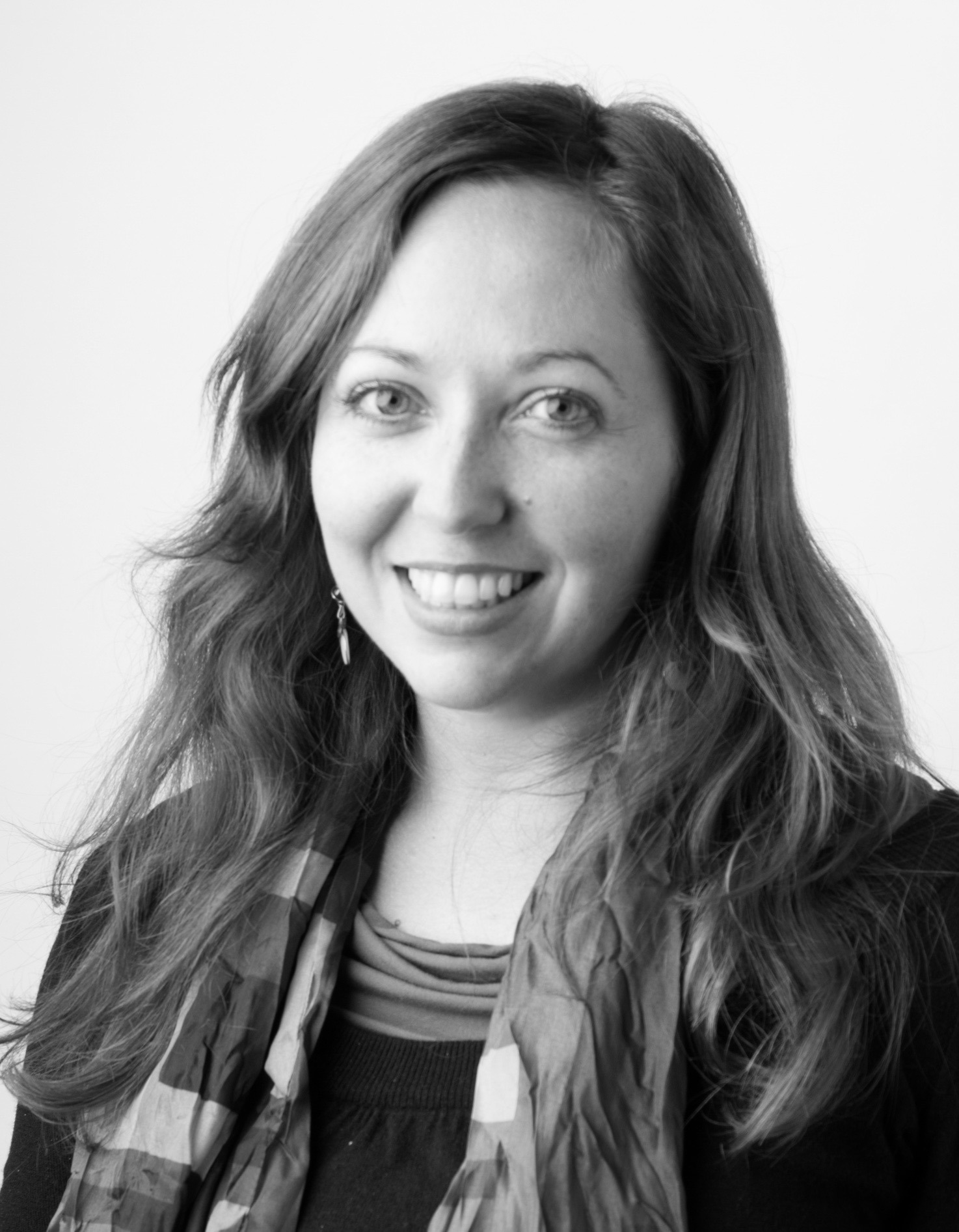}}]{Rikky Muller} (Senior Member, IEEE) is an Associate Professor of Electrical Engineering and Computer Sciences at UC Berkeley where she holds the S. Shankar Sastry Professorship in Emerging Technologies. She received her B.S. and M.Eng. degrees from MIT and her Ph.D. from UC Berkeley all in Electrical Engineering and Computer Science. She is currently a Co-director of the Berkeley Wireless Research Center (BWRC), a Core Member of the Center for Neural Engineering and Prostheses (CNEP) and an Investigator at the Chan-Zuckerberg Biohub. Dr. Muller was previously an IC designer at Analog Devices, and was the co-founder of Cortera Neurotechnologies, Inc. an acquired medical device company focused on closed-loop deep brain stimulation technology. Her research group focuses on emerging implantable and wearable medical devices and in developing low-power, wireless microelectronic and integrated systems for neurological applications. She was named one of MIT Tech Review’s 35 global innovators under 35 (TR35) and Boston MedTech’s 40 healthcare innovators under 40. She is the recipient of the National Academy of Engineering Gilbreth Lectureship, the NSF CAREER Award, the Keysight Early Career Professorship, the McKnight Technological Innovations in Neuroscience Award, and the IEEE Solid-State Circuits Society New Frontier Award. She is a Bakar Fellow, a Hellman Fellow, and a Distinguished Lecturer for the Solid-State Circuits Society.
\end{IEEEbiography}

\end{document}